\documentclass[a4paper,twoside,10pt]{article}
\usepackage[authoryear, round]{natbib}

\usepackage{graphicx, publaob}

\def\papertype{Progress Report}
\begin{document}

% Title
\title{Deep imaging with Milankovi{\'c} telescope: Linking merger history to kinematics of elliptical galaxies}

% Authors
   \authors{Ivana Ebrov{\'a}\,$^{1,*}$,
            Michal B{\'i}lek\,$^{1,2,3}$, 
            Ana Lalovi{\'c}\,$^{4}$,
            Mustafa~K. Y{\i}ld{\i}z\,$^{5,6}$,
            }
    \authors{Pierre-Alain Duc\,$^{7}$, Martin Ma\v{s}ek\,$^{1}$, \lowercase{and} Michael Prouza\,$^{1}$}

%\thanks{\email{ebrova.ivana@gmail.com}

% Addresses and e-mails
\address{$^1$FZU -- Institute of Physics of the Czech Academy of Sciences, Na Slovance 1999/2, 182 00 Prague, Czechia}
\vskip-3mm
\address{$^*$E-mail: ebrova.ivana@gmail.com}
\vskip-3mm
\address{$^2$LERMA, Observatoire de Paris, CNRS, PSL Univ., Sorbonne Univ., 75014 Paris, France}
\vskip-3mm
\address{$^3$Collège de France, 11 place Marcelin Berthelot, 75005 Paris, France}
\vskip-3mm
\address{$^4$Astronomical Observatory, Volgina 7, 11060 Belgrade, Serbia}
\vskip-3mm
\address{$^5$Astronomy and Space Sciences Department, Science Faculty, Erciyes University, Kayseri, 38039 Türkiye}
\vskip-3mm
\address{$^6$Erciyes University, Astronomy and Space Sciences Observatory Applied and Research Center (UZAYB\.{I}MER), 38039, Kayseri, Türkiye}
\vskip-3mm
\address{$^7$Universit{\'e} de Strasbourg, CNRS, Observatoire astronomique de Strasbourg (ObAS), UMR 7550, 67000 Strasbourg, France}

% Running titles
\markboth{\runningfont Linking merger history to kinematics of elliptical galaxies}
{\runningfont I. EBROV{\'A} \runningit{et al.}}

\vskip-1mm

\abstract{
Kinematical and morphological features observed in early-type galaxies provide valuable insights into the evolution of their hosts. We studied the origin of prolate rotation (i.e., rotation around the long axis) in Illustris large-scale cosmological hydrodynamical simulations. We found that basically all the simulated massive prolate rotators were created in relatively recent major mergers of galaxies. Such mergers are expected to produce tidal features such as tails, shells, asymmetric stellar halos. 

We investigated deep optical images of prolate rotators, including newly obtained Milankovi{\'c} data, revealing signs of galaxy interaction in all of them. This correlation proves to be statistically very significant when compared with a general sample of early-type galaxies from the MATLAS deep imaging survey. 

In an ongoing project, we use Milankovi{\'c} to assemble deep images of the complete sample of all known nearby massive prolate rotators. Additionally, we searched these data for asteroids to improve the accuracy of trajectories and even discover one previously unknown main-belt asteroid.

The most frequent tidal features among the prolate rotators happen to be shells. We developed methods to calculate the probable time of the merger from optical images. This will allow us to compare the merger history of the sample with predictions from Illustris. Our plan is to expand these methods to even larger samples of shell galaxies supplied by upcoming large surveys like LSST at Rubin Observatory. This will provide an unprecedented amount of statistically significant data on the recent merger history of our Universe and allow extensive investigation of the impact of mergers to a wide range of other astrophysical phenomena.
}

\vskip-.5cm

%===========================
%===========================
\section{ INTRODUCTION }

The closer we look at a galaxy, the more remarkable characteristics we see. 
Integral-field spectroscopy of inner parts reveals kinematical peculiarities, while deep imaging can uncover various tidal features in the outskirts.
Even many elliptical galaxies that previously seemed featureless have become highly attractive objects to study.
Our work explores the connections between the kinematical and morphological attributes in early-type galaxies (ETGs). 
To unveil these connections, we combine and utilize data and findings from large-scale cosmological simulations on the theoretical side, and integral-field spectroscopy as well as ultra deep imaging on the observational side. 

\vskip-.5mm

%---------------------
\begin{figure} [!b]
\centering
\includegraphics[width=0.92\textwidth,keepaspectratio=true]{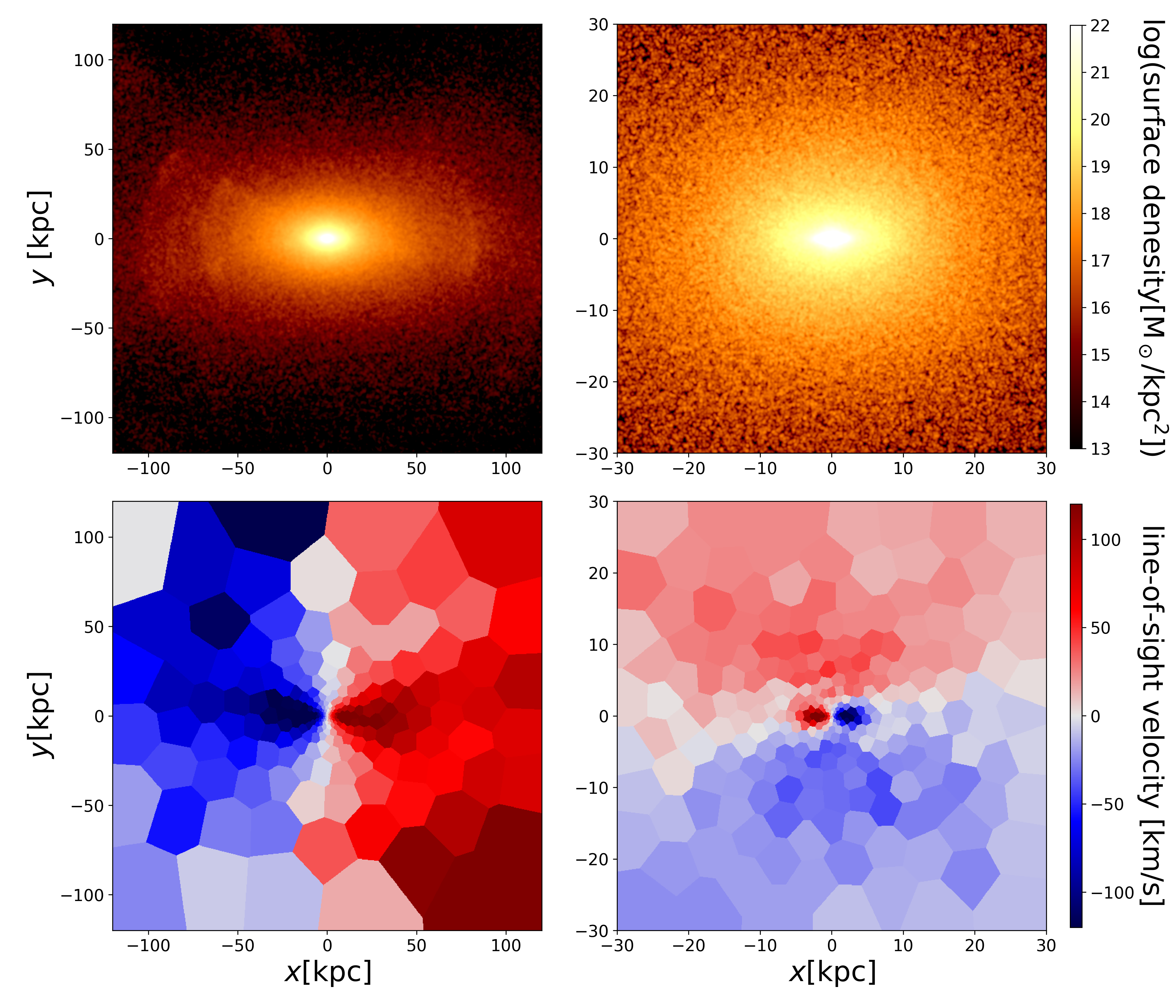}
\vspace{-5mm}
\caption{Surface density (top row) and kinematics (bottom row) of stellar particles of two galaxies from the Illustris simulation. Left column: galaxy with a normal disky rotation and with stellar shells visible in the top panel. Right column: galaxy with prolate rotation and a kinematically distinct core. }
\label{fig1}
\vspace{-1mm}
\end{figure}
%---------------------

%===========================
%===========================
\section{ RESULTS }

%===========================
\subsection{ Simulations }

In \cite{illkdc} and \cite{illprol}, we studied galaxies with kinematical peculiarities in the Illustris project -- hydrodynamic cosmological simulations \citep{vog14illpreintro,nel15illpub}. 
We examined the formation and merger histories of selected galaxies drawn from a global sample of 7697 Illustris galaxies with more than $10^4$ stellar particles (i.e., LMC-like and heavier) in the last snapshot of the Illustris-1 run. 
By visually inspecting kinematic maps, we identified 134 galaxies with kinematically distinct cores (KDCs) and automatically selected 59 galaxies with prolate rotation.

Fig.\ref{fig1} shows two galaxies from Illustris.
On the left, an ETG that maintained normal disky rotation, even though the galaxy suffered a relatively recent merger, as can be inferred from the presence of stellar shells in the surface density map. 
On the right, an ETG with a kinematically distinct core (KDC) and prolate rotation (aalso known as “minor-axis rotation”). 
Prolate rotators exhibit a substantial misalignment between the photometric and kinematic axes;  in other words, the galaxy appears to be rotating predominantly around its major morphological axis. 
The galaxy on the right had both kinematic features created in a major 1:1 merger 6.1 Gyr before the end of the simulation. 

%---------------------
\begin{figure} [!b]
\centering
\includegraphics[width=\textwidth,keepaspectratio=true]{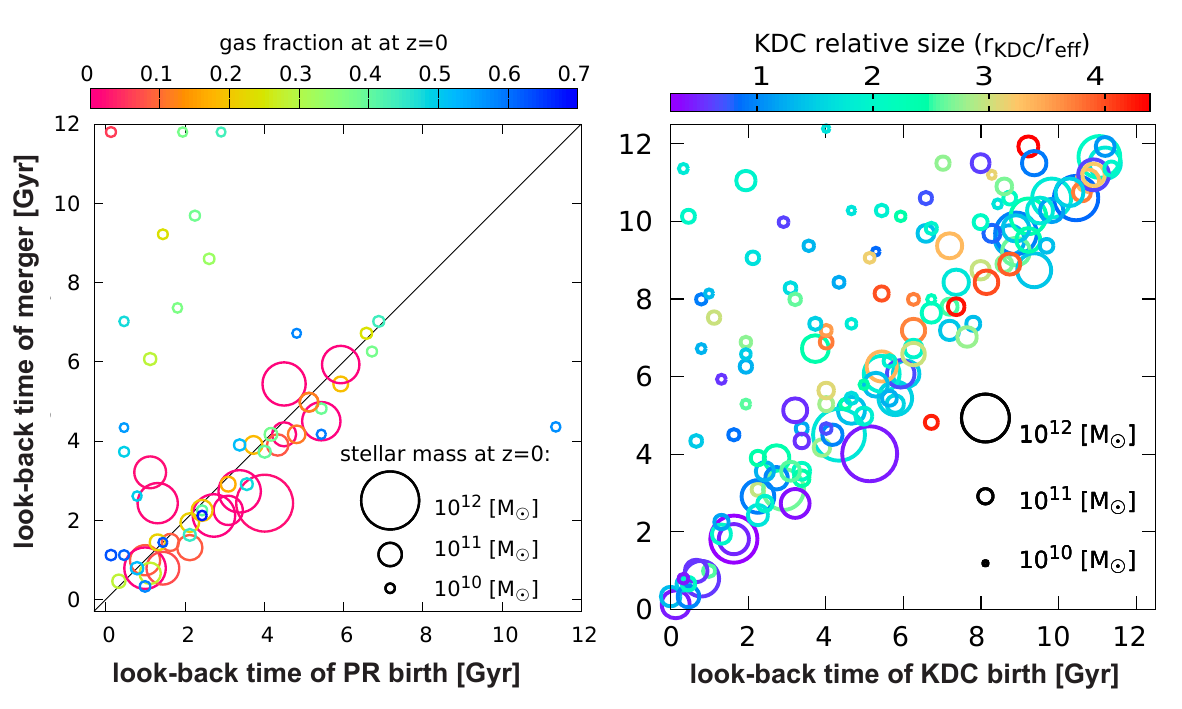}
\vspace{-8mm}
\caption{Correlation of the birth of prolate rotation (left) and KDCs (right) with the time of the merger experienced by the host galaxy in the Illustris simulation. 
For prolate rotators, the merger time represents the last significant merger, while for the KDC hosts, it indicates the time of the merger closest to the KDC birth. Circle areas are proportional to the host stellar mass at the end of the simulation. }
\label{fig2}
\vspace{-1mm}
\end{figure}
%---------------------

While both, prolate rotation and KDCs, can emerge from mergers, Illustris data indicate systematic differences between those two.
Prolate rotation is more common among massive ETGs, consistent with observations. 
In contrast, KDCs display no clear dependence on the host mass and environment.
We specifically examined the role of galaxy mergers in creating both kinematic features, see Fig.\ref{fig2}.
KDCs more often have other origins, and if their origins are associated with mergers, the mergers can be minor or ancient.
Other KDCs are induced by galaxy fly-bys or without an apparent cause.
Moreover, KDCs can be long-lasting features and survive subsequent significant mergers.

On the other hand, basically all massive prolate rotators were created in major mergers (at least 1:5) during the last 6 Gyr of the Illustris simulation. 
Such mergers are expected to produce tidal features that should be, in the majority of cases, visible in current deep imaging surveys.
Based on the Illustris data, we predicted that the frequency of tidal features in host galaxies should be higher for prolate rotation than for KDCs.

%---------------------
%\begin{figure} [!ht]
\begin{figure} [!b]
\vspace{-1mm}
\centering
\includegraphics[width=0.98\textwidth,keepaspectratio=true]{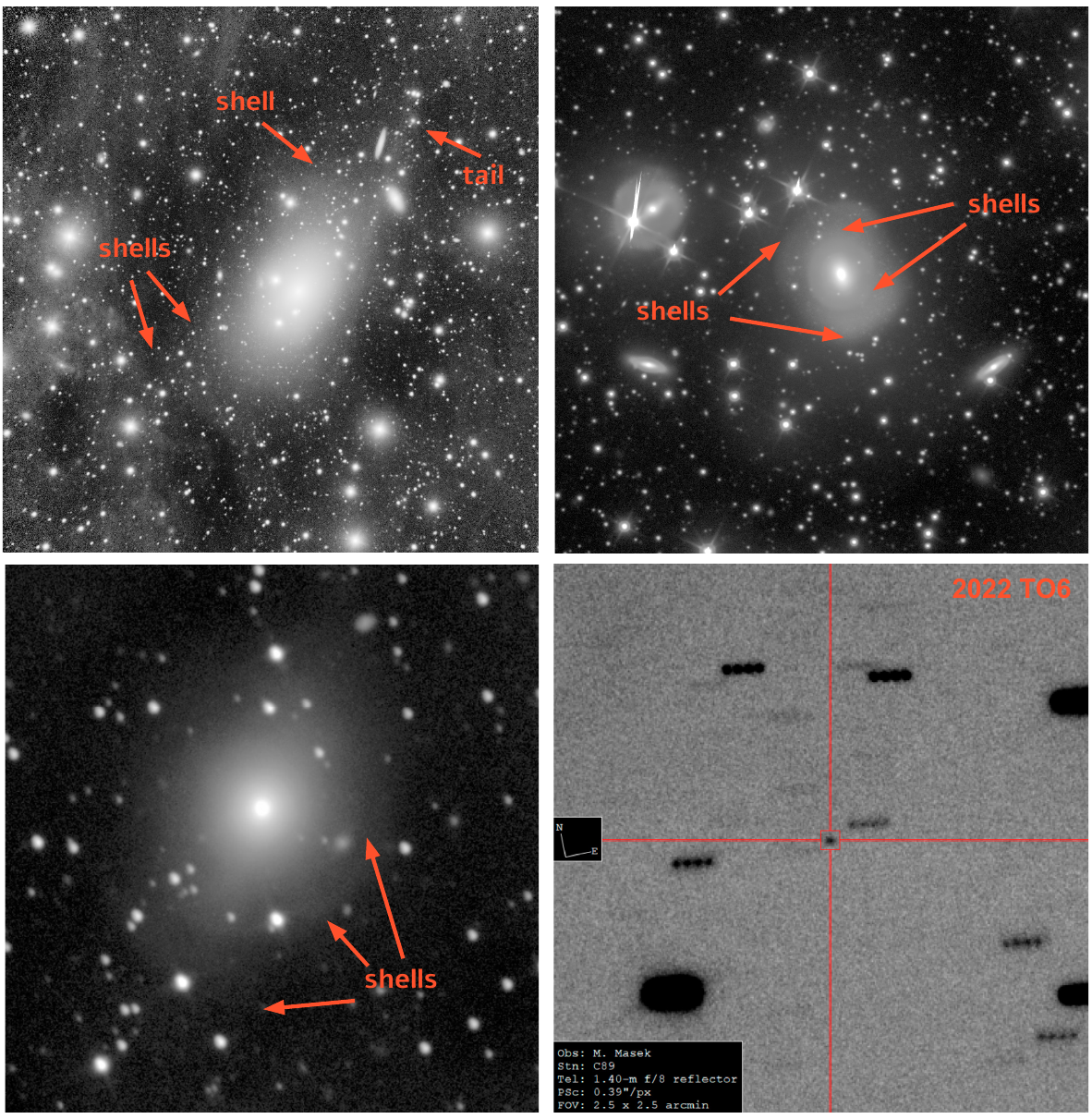}
\vspace{-3mm}
\caption{Images from the Milankovi{\'c} telescope: recent deep images of three prolate rotators and, in the bottom right panel, the newly discovered asteroid 2022 TO6.}
\label{fig3}
\vspace{-1mm}
\end{figure}
%--------------------- 

%===========================
\subsection{ Observations }

In \cite{probs} we examined 19 observed prolate rotators with available deep optical images and found morphological signs of galaxy interaction in all of them, which proves to be a statistically very significant correlation when compared with a general sample of ETGs in MATLAS -- a deep imaging survey \citep{matlas20}.

In our current project, we use the Serbian 1.4m Milankovi{\'c} telescope at the Astronomical Station Vidojevica to assemble deep optical images of the complete sample of all known nearby massive prolate rotators. 
Between Feb 2021 and Oct 2023 we observed 5 out of 8 additional prolate rotators, each at least 5.5\,h integrated on-source exposure time in the $L$-band. 
Fig.\ref{fig3} shows preliminary processed deep images of three of these prolate rotators. All show signs of galaxy interactions.

%===========================
\subsection{ Small Solar System bodies }

We search and measure small Solar System bodies as a by-product of deep imaging of galaxies with the Milankovi{\'c} telescope. 
We use Tycho-Tracker to explore the field of view of the images of prolate rotators and other galaxies. 
In some cases, we performed follow-up and dedicated observations.
So far, we have significantly improved trajectory measurements of more than 50 objects of 17\,--\,22\,mag and even one asteroid as faint as 23.2\,mag in stacked images. 

However, the most remarkable is the discovery of a previously unknown main-belt asteroid that received a temporary designation 2022\,TO6, see the bottom right panel of Fig.\ref{fig3} -- the stacked image of 4$\times$300\,s exposures, each centered on the asteroid before stacking.
As far as we know, this is the first and currently the sole asteroid discovery at the Astronomical Station Vidojevica.

%===========================
\subsection{ Merger histories }

The most frequent tidal features among the prolate rotators happen to be stellar shells.
Therefore, we can estimate the timing of mergers for a large portion of the sample and compare it with the predictions of the merger history of prolate rotators in the Illustris simulation.
 
We developed the ‘shell identification method’ in \cite{bil13} and \cite{bil14}. So far, we have applied it to several special cases of shell galaxies to explore the host gravitational potential and derive the time of the galaxy mergers undergone by the hosts (\citealp{bil14, e20sg, bil22}; see also \citealp{bil15}).

There are hundreds of known shell galaxies, even more hidden in current data, and much more will be observed in the next few years in upcoming large deep surveys like the Large Survey of Space and Time (LSST) at the Vera C. Rubin Observatory. 
We are developing tools to extract fairly accurate estimates of the merger times for large samples. 
This will transform shell galaxies from a position of curiosity to utility, allowing statistical applications using the merger data on thousands of shell galaxies.
This will provide an unprecedented amount of data on the recent merger history of our Universe and allow extensive investigation of the impact of mergers on a wide range of other astrophysical phenomena such as star formation, stellar dynamics, active galactic nuclei, transient events, and more.

%\vskip0.5cm

%===========================
%===========================
\section{ CONCLUSIONS }

We investigated the link between kinematical and morphological features in early-type galaxies through simulations as well as observations. 
In Illustris, we found that while the origin of kinematically distinct cores is partially associated with mergers of galaxies, basically all massive prolate rotators were created in relatively recent major mergers.
Such mergers are expected to leave tidal features that should be detectable today in sufficiently deep images.

In our analysis of available observational data, we found an overabundance of morphological signs of galaxy interactions in prolate rotators. With the help of the Milankovi{\'c} telescope, we are assembling deep optical images of the complete sample of all known nearby massive prolate rotators. 
Initial data processing confirms the high statistical significance of previous findings, showcasing Milankovi{\'c} as a valuable tool for studying low-surface-brightness tidal features. 
Additionally, the same data can be used to search for small Solar System bodies. 
We significantly improved the accuracy of trajectories for more than 50 such objects and even discovered a previously unknown main-belt asteroid 2022\,TO6.

The most frequent tidal features among the prolate rotators are stellar shells that can be used to constrain the time since the merger. 
That will enable us to compare merger histories of prolate rotators with the Illustris predictions.
Our plan is to extend such analyses to hundreds, potentially thousands, of shell galaxies. 
This will significantly deepen our understanding of the impact of mergers on various astrophysical phenomena.

%===========================
%===========================
{\vskip 3mm}
\centerline{\bf Acknowledgements}
{\vskip 3mm}

\noindent 
This project has received funding from the European Union's Horizon Europe Research and Innovation program under the Marie Skłodowska-Curie grant agreement No. 101067618.
We acknowledge support by the Astronomical Station Vidojevica and funding from the Ministry of science, technological development and innovation of the Republic of Serbia, contract No. 451-03-47/2023-01/200002 and by the EC through project BELISSIMA (call FP7-REGPOT-2010-5, No. 256772). 
IE acknowledges the support from the Polish National Science Centre under the grant 2017/26/D/ST9/00449.
%\vskip-.5cm

%%%%%%%%%%%%%%%%%%%% REFERENCES %%%%%%%%%%%%%%%%

%\references

%\bibliographystyle{aasjournal}
%\bibliographystyle{mnras,abbrvnat}
%\bibliographystyle{aa}
\bibliographystyle{abbrvnat}
\bibliography{prolrot.bib}

%\endreferences

%%%%%%%%%%%%%%%%%%%%%%%%%%%%%%%%%%%%%%%%%%%%

\end{document}